\documentclass{Rinton-P9x6}
\usepackage{epsf}
\begin{document}

\title{Casimir interaction between cylinders}

\author{Francisco D. Mazzitelli}

\address{Departamento de F\'\i sica {\it J.J. Giambiagi}, Facultad de Ciencias Exactas y Naturales,
Universidad de Buenos Aires - Ciudad Universitaria, Pabell\' on I,
(1428) Buenos Aires, Argentina.
}

\maketitle

\abstracts{We compute the Casimir interaction energy between  two
perfectly conducting, concentric cylinders, using the mode-by-mode
summation technique. Then we compare it with the approximate
results obtained using the proximity theorem and a semiclassical
approximation based on classical periodic orbits. We show that the
proximity theorem with a particular choice for the effective area
coincides with the semiclassical approximation and reproduces the
exact result far beyond its expected range of validity. We also
compute the force between slightly eccentric cylinders and discuss
the advantages of using a cylindrical geometry to measure the
Casimir force.}

\section{Introduction}

Many of the recent experiments that measured the Casimir force
between conductors involved the interaction between a plane and a
sphere, \cite{exp} with only one exception  where the force
between parallel plates was measured.\cite{roberto} The
plane-sphere configuration is more appealing from an experimental
point of view, mainly because the alignment problem that
complicates the measurement for the parallel plates is not
present. However, from a theoretical point of view, the
inconvenience is that there is no exact evaluation of the force
between a sphere and a plane. The comparison between theory and
experiment is done with the ``proximity"  approximation for the
force.

It is then of interest to explore other geometries for which the
Casimir force can be computed exactly. This would be useful both
for testing the validity of the approximations and, eventually,
for proposing new experiments. In what follows we will be
concerned with the Casimir interaction between two long,
perfectly conducting cylindrical shells of radii $a$ and $b$ and
length $L\gg a,b$.

\section{Exact Casimir energy}\label{sex}

In order to define properly the Casimir energy we introduce a
cutoff $\sigma$ as follows
\begin{equation}
E_{C}(\sigma)={\hbar\over 2}\sum_{p=n,m,k_z}(e^{-\sigma w_p}
w_p-e^{-\sigma \tilde w_p} \tilde w_p) \; . \label{ecasreg}
\end{equation}
The exact Casimir energy $E_{C}$ is the limit of $E_{C}(\sigma)$
as $\sigma\rightarrow 0$. In our definition we take $w_p$ as the
eigenfrequencies of the electromagnetic field satisfying perfect
conductor boundary conditions at $r=a, r=b$ and $r=R$, and $\tilde
w_p$ are those corresponding to the boundary conditions at $r=R_1,
r=R_2$ and $r=R$, in the limit $R>R_2>R_1\gg a > b$. The
parameters $R_1, R_2$ and $R$ define the reference vacuum.

In cylindrical coordinates, the eigenfunctions are of the form
\begin{equation}
h_{n k_z}(t,r,\theta,z)=e^{(-iw_{n k_z}t+ik_z
z+in\theta)}R_n(\lambda r) \; , \label{eigenf}
\end{equation}
where the function $R_n$ is a combination of Bessel functions
satisfying the perfect conductor boundary conditions. These
boundary conditions define the possible values of the constant
$\lambda$. The eigenfrequencies are $w_{n k_z}=c
\sqrt{k_z^2+\lambda^2}$. The frequencies of the TE modes are
defined by
\begin{eqnarray}
F_n^{TE}(z,a,b,R)& \equiv &J_n(z a)[J_n(z a)N_n(z b)-J_n(z b)
N_n(z
a)]\nonumber\\
&\times &[J_n(z b)N_n(z R)-J_n(z R) N_n(z b)]=0 \; .
\end{eqnarray}
The frequencies of the TM modes involve derivatives of the Bessel
functions, and one can define an analogous function $F_n^{TM}$. In
the set of quantum numbers $p=(n,m,k_z)$ appearing in Eq.
(\ref{ecasreg}), $m$ denotes the different solutions
$\lambda_{nm}$ of both $F_n^{TE}(z,a,b,R)=0$ and
$F_n^{TM}(z,a,b,R)=0$ .

From Cauchy's theorem it follows that
\begin{equation}
{1\over 2\pi i} \int_C \,dz \;  z \; e^{-\sigma z} {d\over dz}\ln
f(z)=\sum_i x_i \; e^{-\sigma x_i} \; ,
\end{equation}
where $f(z)$ is an analytic function within the closed contour
$C$, with simple zeros at $x_1,x_2,...,$ within $C$. We use this
result to replace the sum over $m$ in Eq.(\ref{ecasreg}) by a
contour integral in the complex plane. \cite{piro} We find
\cite{javier}
\begin{equation}
E_C (\sigma)={L\hbar c\over 4\pi i}\int_{-\infty}^{\infty}
{dk_z\over 2\pi} \sum_n\int_C dz \sqrt{k_z^2+z^2} e^{-\sigma c
\sqrt{k_z^2+z^2}}{d\over dz}\ln F^{EM}_{n}(z,a,b) \; ,
\end{equation}
where
\begin{equation}
F^{EM}_{n}(z,a,b)= \lim_{R_{1},R_{2},R \rightarrow\infty}
{F_n^{TE}(z,a,b,R) F_n^{TM}(z,a,b,R)\over F_n^{TE}(z,R_1,R_2,R)
F_n^{TM}(z,R_1,R_2,R)} \; .
\end{equation}
An adequate contour for the integration in the complex plane is a
circular segment $C_{\Gamma}$ of radius $\Gamma$  and two straight
line segments forming an angle $\phi$ and $\pi -\phi$ with respect
to the imaginary axis. The nonzero angle $\phi$ is needed to show
that the contribution of $C_{\Gamma}$ vanishes in the limit
$\Gamma\rightarrow \infty$ when $\sigma > 0$.

It proves to be convenient to compute the difference between the
energy of the system of two concentric cylinders and the energy of
two isolated cylinders of radii $a$ and $b$
\begin{equation}
E_{12}(\sigma)=E_C(\sigma) - E_1(\sigma,a)-E_1( \sigma,b) \; .
\end{equation}
The divergences in $E_{C}(\sigma)$ are cancelled out by those of
$E_1(\sigma,a)$ and $E_1(\sigma,b)$. Therefore, to compute
$E_{12}(\sigma)$ we can set $\phi=0$ and $\sigma =0$. The contour
integral reduces to an integral on the imaginary axis. After some
steps we find \cite{javier}
\begin{equation}
E_{12}= -{L\hbar c \over 2\pi a^2}\int_{-\infty}^{\infty}
{dk_z\over 2\pi} \sum_n Im \left\{\int_0^{\infty} dy
\sqrt{k_z^2-y^2} {d\over dy}\ln F_{n}(y,\alpha)\right\} \; ,
\label{xx}
\end{equation}
where $\alpha=b/a$ and
\begin{equation}
F_{n}(y,\alpha)=\left[1-{I_n(y)K_n(\alpha y)\over I_n(\alpha
y)K_n(y)}\right] \left[1-{I'_n(y)K'_n(\alpha y) \over I'_n(\alpha
y)K'_n(y)}\right]\,\,\,\,  . \label{bessel}
\end{equation}
Eqs. (\ref{xx}) and (\ref{bessel}) allow a simple numerical
evaluation of $E_{12}$. The exact energy for the two concentric
cylinders is the sum of $E_{12}$ and the Casimir energies for
single cylinders of radii $a$ and $b$ \cite{nesterenko}
\begin{equation}
E_{C} = E_{12} - 0.01356 \; (\frac{1}{a^2}+\frac{1}{b^2}) \; L\;
\hbar \;
 c \; .
\label{exfin}
\end{equation}

\section{Proximity and semiclassical approximations}

The Casimir energy per unit area for parallel plates separated by
a distance $l$ is given by $E_{pp}(l)= -{\pi^2 \hbar c \over 720
l^3}$. In the proximity approximation, the interaction energy
between two conductors that form a curved gap of variable width
$z$ can be computed as \cite{blocki}
\begin{equation}
E_{I}=\int_\Sigma E_{pp}(z) d\sigma \,\,\, , \label{pp}
\end{equation}
where $\Sigma$ is one of the two surfaces that define the gap. It
is clear that in the above approximation the non-parallelism of
the area elements is not taken into account. Moreover, the result
is different if the other surface is chosen to perform the
calculation. These corrections are expected to be small for low
curvature and close surfaces.

For the concentric cylinders, the proximity approximation gives
\begin{equation}
E_{I}(b-a)= -{\pi^2 \hbar c \over{720}}{A_{eff}\over
(b-a)^3}\,\,\, . \label{cc}
\end{equation}
As mentioned before, there is an ambiguity in the choice of the
effective area $A_{eff}$. A calculation based on the inner
cylinder gives $A_{eff}=2\pi a L$, while for a calculation based
on the outer one gives $A_{eff}=2\pi b L$. The difference is of
course harmless in the limit $\alpha\rightarrow 1$. In this limit,
the exact result for $E_{12}$ coincides with the proximity
approximation. Indeed, using the uniform expansion of Bessel
functions in Eqs. (\ref{xx}) and (\ref{bessel}), one can show that
to leading order in $\alpha - 1$,   $E_{12}\approx E_I(b-a)$.

For larger values of $\alpha$ one expects the exact result to
differ significantly from the proximity approximation. We have
computed numerically the interaction energy $E_{12}$, and compared
it with the proximity approximation, using different choices for
the effective area. In particular, we parametrized the effective
area as $A_{eff}=2\pi L a^pb^{1-p}$, and found that the value
$p=1/2$ provides the best fit for the numerical data. For this
value of $p$, the effective area is the geometric mean of the
areas of both cylinders. As shown in Fig. 1, the proximity
approximation with this particular choice for the area reproduces
the exact results far beyond its expected range of validity:
\cite{javier} the discrepancy for the pressure on the inner
cylinder due to the presence of the outer one is less than $10\%$
for $1<\alpha <4$. The choice of the geometric mean is crucial
for this agreement.
\begin{figure}[b]
\begin{center}
\epsfxsize=20pc 
\epsfbox{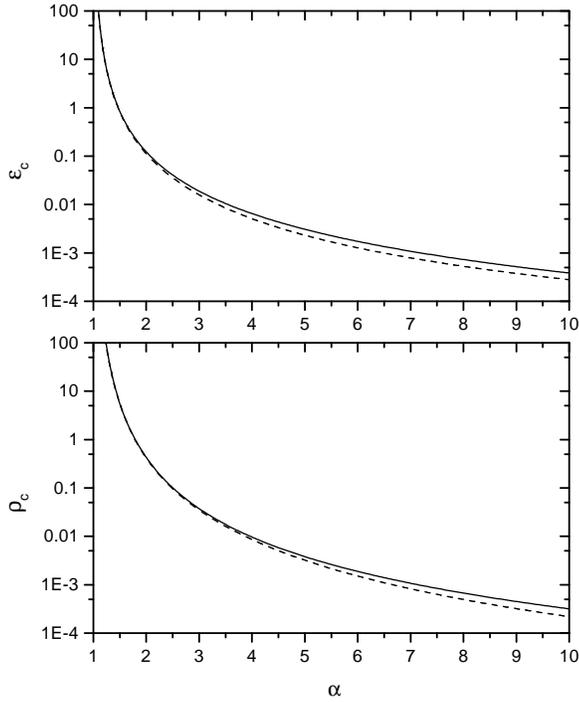} 
\end{center}
\caption{Dimensionless Casimir interaction energy (upper panel)
and pressure (lower panel), as a function of $\alpha = b/a$. In
both panels the dashed line corresponds to the exact result and
the full line to the $p=1/2$ proximity result.
 \label{fig:res}}
\end{figure}

Let us now discuss briefly the semiclassical approximation for the
Casimir energy, which can be written as
\begin{equation}\label{C}
 E_{C} = \int_{0}^{\infty} \frac{1}{2} \;  E \;   \rho_{osc} (E) \; dE \; ,
\end{equation}
where $\rho_{osc}(E)$ is the difference between the spectral
density of electromagnetic modes in the presence of conductors and
the spectral density in vacuum. Periodic orbit theory relates
$\rho_{osc}$ of a given Hamiltonian to the periodic orbits in the
corresponding classical system. \cite{javier,spruch,new} To
leading order in $\hbar$,
\begin{equation}
\rho^{osc} (E)= \frac{1}{\hbar^\nu} \sum_t \; A_{t}(E) \;  \sin (
S_{t}(E)/ \hbar
 \; + \mu_{t} ) \; ,
\end{equation}
where the sum runs over periodic orbits labeled by $t$, and
$S_{t}$ is the classical action of the periodic orbit  $t$. For
photons of energy $E$, $S_t(E)=E L_t/c$, with $L_{t}$ the length
of the periodic orbit. The phase $\mu_t$ is the so-called Maslov
index. The exponent $\nu$ and the amplitudes $A$ depend on the
type of the periodic orbit.

The periodic orbits in the region between very long concentric
cylinders are contained in planes perpendicular to the axis of the
cylinders. There are two different classes of orbits. The type-I
orbits do not touch the inner cylinder, and are polygons that may
be uniquely labeled by two integers $(v, w)$, where $v$ is the
number of bounces in the outer cylinder and $w$ the winding
number around the center. Type-II trajectories  do touch both
cylinders, and are also labeled by $(v, w)$. We have have
computed the Casimir interaction energy using this semiclassical
approximation. \cite{javier} The final result is dominated by the
self-retracing type II periodic orbit $(w=0)$ and its
repetitions. {\it Moreover, it coincides with the proximity
approximation if the  effective area is taken as the geometric
mean of the areas of both cylinders}.

\section{Eccentric cylinders: an experimental proposal}

Up to now we considered concentric cylinders. Obviously, the force
between them vanishes in this case. Let us now consider two
slightly eccentric cylinders of radii $a$ and $b$. We will mainly
focus on the particular case $a\simeq b$, since the Casimir
interaction between cylinders is then stronger. The distance
between the axis of the cylinders will be denoted by $\epsilon$
(see Fig. 2). In order to evaluate the Casimir energy for this
configuration, we will use the proximity approximation. This is
partially justified by our previous results.
\begin{figure}[t]
\begin{center}
\epsfxsize=15pc 
\epsfbox{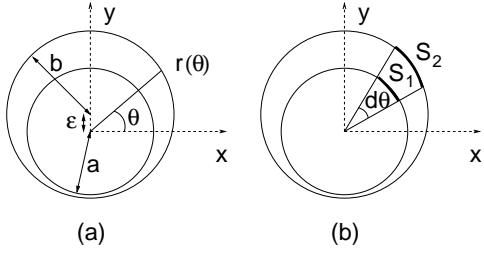} 
\end{center}
\caption{Two eccentric cylinders.  (a) An inner cylinder of
radius $a$ and a hollow cylinder of radius $b$, with the origin
of coordinates on the axis of the inner cylinder, and  distance
$\epsilon$ between the two axis. (b) The effective area for the
application of the proximity approximation, as the geometric mean
of $S_1$ and $S_2$:  $dA_{\rm eff}(\theta) =\sqrt{S_1 S_2}$
\label{fig:ecc}}
\end{figure}
The interaction energy between cylinders is
\begin{equation}
E_I \simeq -{\pi^2 \hbar c   \over 720} \int_0^{2\pi} \frac{d
A_{\rm eff}(\theta)}{[ r(\theta ) - a]^3}, \label{prox}
\end{equation}
where  $r(\theta )$ is the distance of a point of the external
cylinder to the axis of the inner one and $dA_{eff}(\theta)$ is
the geometric mean of two small adjacent areas on both cylinders.
From Fig.2 one can check that $r(\theta )= \sqrt{b^2 - \epsilon^2
\cos^2\theta} + \epsilon \sin\theta$ and $dA_{eff}=
L\sqrt{ab+\epsilon a\sin\theta}d\theta$. Taking the derivative of
the energy with respect to $\epsilon$ we obtain the force on the
outer  cylinder due to the inner one. Since we are considering
$a\simeq b$, we will always have $\epsilon/b \ll 1$. Thus to the
lowest non trivial order we obtain \cite{prep}
\begin{equation}
F_y= -{\pi^2 \hbar c L a\over 240 b^4} \int_0^{2\pi} {d\theta
\sin\theta\over \left[ 1 + {\epsilon\over b} \sin\theta
 - {a\over b}\right]^4}\approx F_0 {\left(\tilde\epsilon
 + {{\tilde\epsilon}^3 \over 4}
\right) \over{\left[1 - \tilde\epsilon^2\right]^{7\over{2}}}}
.\label{result1}
\end{equation}
where $\tilde\epsilon = \epsilon/(b-a)$ and $F_0=\pi^3 \hbar c L a
/ 60 (b-a)^4$. It is worth noting that closest surfaces are
attracted together, and that the force  only vanishes when the
cylinders are exactly concentric. {\it The equilibrium position is
unstable}. In the particular case in which $\tilde\epsilon \ll 1$,
the force is linear in the distance between the axis of the
cylinders $ F_y \approx \tilde\epsilon F_0 $. This corresponds to
an inverted harmonic oscillator, and explicitly shows the
instability . In the opposite case, when $\tilde\epsilon
\rightarrow 1$,  the force scales like $d^{-7/2}$, where
$d=b-a-\epsilon$ is the minimum distance between cylinders.

We now discuss a possible experimental arrangement that could be
used to measure the Casimir force between cylinders. \cite{prep}
We consider the almost coaxial configuration $\tilde\epsilon\ll
1$. The external cylinder could be mounted on a resonator of
effective mass $M$ and natural frequency $\omega_0$. The presence
of the inner cylinder renormalizes the frequency of the resonator.
Assuming a small frequency shift we obtain ${\Delta \omega \over
\omega_0}=-{F_0\over 2(b-a)M \omega_0^2}$. For typical values of
the parameters it is simple to reach a frequency shift of 0.1\%.
From an experimental point of view, this configuration has some
advantages over the parallel plates configuration. On the one
hand, the alignment procedure is easier for the cylinders than for
the plates (see below). On the other hand, when there is no
residual charge in the inner cylinder, the system remains neutral
and screened by the external one from outer noises, interferences,
and from residual charges in the outer cylinder. The expected
gravitational force is obviously null. When the inner cylinder has
a residual charge, there will be a small potential difference
between the cylinders, and the coaxial configuration will be
electrostatically unstable. To avoid it, one could start the
experiment by putting both cylinders in contact. Ideally, the
charge of the inner cylinder would flow to the outer one,
minimizing the effect if residual electrostatic forces.
Alternatively, one could use the electrostatic instability to
improve the alignment: if a time dependent differential potential
is applied between cylinders (as in the null experiments to test
the exactness of the electrostatic inverse square law),
parallelism would be optimized when the value of the force is
minimum.

To summarize, we have shown that the interaction between
cylindrical shells is both of theoretical and experimental
interest for the study of Casimir forces. On the theoretical
side, it is a simple but non trivial example for testing the
validity of different approximations to the Casimir energy. On
the experimental side, it is a promising geometry for measuring
the Casimir force.

\section*{Acknowledgments}
The work reported here has been done in collaboration with the
authors of Refs. [4] and [9]. I thank all of them.  This work was
supported by Universidad de Buenos Aires, Conicet, and Agencia
Nacional de Promoci\'on Cient\'\i fica y Tecnol\'ogica, Argentina.


\begin{thebibliography}{99}

\bibitem{exp} S. K. Lamoreaux, \Journal{\PRL}{78}{5}{1997};
U. Mohideen and A. Roy, \Journal{\PRL}{81}{4549} {1998}; H.B.
Chan {\it et al.}, \Journal{\PRL}{87} {1801}{2001}

\bibitem{roberto} G. Bressi, G. Carugno, R. Onofrio and G. Ruoso,
\Journal{\PRL} {88}{041804} {2002}.


\bibitem{piro}V.V. Nesterenko and I.G. Pirozhenko, \Journal{\PRD}{57}
{1284} {1997}.

\bibitem{javier} F.D. Mazzitelli, M.J. Sanchez, N. Scoccola and J. Von Stecher,
{\sl Phys. Rev.}A {\bf 67}, 013807 (2003).


\bibitem{nesterenko}L.L. DeRaad, Jr. and K. A. Milton, {\sl Ann. Phys.} {\bf 136}, 229 (1981);
K.A. Milton, A.V. Nesterenko and V.V. Nesterenko,
\Journal{\PRD}{59}{105009} {1999}; P. Gosdzinsky and A. Romeo,
\Journal{\PLB}{441}{265}{1998}.

\bibitem{blocki}
J. Blocki, J. Randrup, W.J. Swiatecki and F. Tsang, {\sl Ann.
Phys.}{\bf 105}, 427 (1977).

\bibitem{spruch} M. Schaden and L. Spruch, {\sl Phys. Rev.} A {\bf 58},
935 (1998).

\bibitem{new} For related works see the contribution of S.A. Fulling in
these Proceedings; see also R.L. Jaffe and A. Scardicchio, 
\Journal{\PRL} {92}{070402} {2004}

\bibitem{prep} D.A. Dalvit, F.C. Lombardo, F.D. Mazzitelli and R. Onofrio, 
{\sl Casimir force between eccentric cylinders}, quant-ph/0406060, to appear
in Europhysics Letters.

\end{thebibliography}
\end{document}